\documentclass[twocolumn,groupedaddress,
superscriptaddress,preprintnumbers,
amsmath,amssymb,
aps,
]{revtex4}
\usepackage{graphicx}
\usepackage{dcolumn}
\usepackage{bm}%
\usepackage[colorlinks=true,citecolor=blue]{hyperref}
\hypersetup{colorlinks=true,citecolor=blue,linkcolor=red,urlcolor=blue}

\usepackage{float}
\usepackage{amsmath}

\usepackage{color}
\definecolor{Green}{RGB}{0,204,102}
\definecolor{Purple}{RGB}{102,0,255}
\definecolor{Blue}{RGB}{51,153,255}
\definecolor{Red}{RGB}{151,010,010}

\makeatletter
\def\@bibdataout@aps{%
\immediate\write\@bibdataout{%
@CONTROL{%
apsrev41Control%
\longbibliography@sw{%
    ,author="08",editor="1",pages="1",title="0",year="1"%
    }{%
    ,author="08",editor="1",pages="1",title="",year="1"%
    }%
  }%
}%
\if@filesw \immediate \write \@auxout {\string \citation {apsrev41Control}}\fi 
}
\makeatother

\begin{document}

\title{Magnetic order tuning of excitons  in the magnetic semiconductor CrCl$_3$ through strain}
\author{Ali Ebrahimian}
\email{ali.ebrahimian@inv.uam.es}
\affiliation{Condensed Matter Physics Center (IFIMAC), Universidad Autónoma de Madrid, 28049, Madrid, Spain}
\affiliation{Departamento de Física de la Materia Condensada, Universidad Autónoma de Madrid, E-28049 Madrid, Spain}
\author{Francisco J. García-Vidal}
\affiliation{Condensed Matter Physics Center (IFIMAC), Universidad Autónoma de Madrid, 28049, Madrid, Spain}
\affiliation{Departamento de Física Teórica de la Materia Condensada, Universidad Autónoma de Madrid, E-28049 Madrid, Spain}
\author{Juan J. Palacios}
\affiliation{Condensed Matter Physics Center (IFIMAC), Universidad Autónoma de Madrid, 28049, Madrid, Spain}
\affiliation{Departamento de Física de la Materia Condensada, Universidad Autónoma de Madrid, E-28049 Madrid, Spain}
\affiliation{Instituto Nicolás Cabrera (INC), Universidad Autónoma de Madrid, E-28049 Madrid, Spain}
\date{\today}

\begin{abstract}
Magnon-exciton coupling provides an unprecedented opportunity for the optical tunability of spin information and, viceversa, for the magnetic control of the optical response.  Few-layered magnets are an ideal platform for the experimental study of this coupling, in part due to the  strong excitonic character of the optical response in (quasi-)two dimensions.   Here, we demonstrate a strong dependence of the excitonic oscillator strength on the magnetic order in a monolayer of chromium trichloride CrCl$_3$.
Solving an effective Bethe–Salpeter equation, we evaluate the changes in the excitonic response as the magnetic order switches from the ferromagnetic state to the antiferromagnetic state when strain is applied. 
Our results reveal an abrupt change in the spatial localization of the excitons across the transition, which translates into a strong shift and a change of the oscillator strength of the excitonic peaks. This suggests the possibility of using strain as a binary switch of the optical response in this material.

\end{abstract}

\maketitle


\section{Introduction}\label{sec:intro}

Due to quantum confinement, poor screening, and the presence of a band gap, two-dimensional (2D) semiconductor magnets offer the unique opportunity to study tightly bound excitons, spin dynamics, and their mutual interactions \cite{dirnberger2023magneto, klein2022control, marques2023interplay}. In addition, external tunability of magnetic order by various means enlarges the playground for a better microscopic understanding of these interactions and paves the way for new magneto-optical applications. 
All this is now possible after the recent discovery of the layered van der Waals magnet CrSBr, which shows magnetic order with potentially long-lived coherent magnons (quantized spin-wave excitations) and a coherent modulation of the electronic structure that results in an outstanding modification of the dominant excitonic excitations \cite{wilson}.

The impact of the magnetization direction on the exciton energies and wavefunctions in bilayer CrSBr has been reported \cite{wilson}, showing that the excitonic transitions in CrSBr can be drastically changed when the magnetic order is changed from layered antiferromagnetic to a field-induced ferromagnetic state. The study of photoluminescence under the variation of an external magnetic field showed a measurable redshift in excitonic emission (of the order of 20 meV) for bilayer and few-layer films of CrSBr. It was  concluded that the coupling of Wannier excitons to interlayer magnetic order is related to the spin-dependent interlayer electron exchange interaction.
More interestingly, the temporal modulation of this canting angle between the magnetization vectors of two antiferromagnetic CrSBr sublattices was further confirmed by time-resolved optical spectroscopy \cite{diederich2023tunable, bae2022exciton, sun2024dipolar}.

Magnon-exciton coupling provides the possibility of spin-wave detection from an optical response within the energy range of excitonic transitions. Despite these burgeoning developments and the potential application of the magnon-exciton coupling, materials displaying these properties are limited, being CrSBr with A-type antiferromagnetism ($T_N$ = 140 K) \cite{lee2021magnetic} and quasi-1D electronic structure the only material that clearly features such a coupling to this date.
Therefore, for our understanding of magnon-exciton interactions, it is fundamentally and technologically significant to extend the range of materials where these interactions manifest.

Motivated by this, herein, we focus on CrCl$_{\rm 3}$ monolayer, which exhibits a ferromagnetic (FM)-antiferromagnetic (AFM) transition at moderate strain fields. In fact,
previous reports have shown that the magnetic anisotropy and magnetic state of monolayer CrCl$_{\rm 3}$ can be tuned by strain and also by electric fields\cite{dupont2021monolayer}. In particular, it has been shown that biaxial strain allows for such a transition. Being able to tune the FM-AFM transition only with strain free up experimental complexity, improving the chances for practical applications. 

Changing the magnetic ground state is expected to have an impact on the excitonic wave functions and the optical response \cite{Han.PhysRevB.110.115418}.
We study this by solving the Bethe–Salpeter equation (BSE) on top of a density functional theory (DFT) calculation, as implemented in XATU\cite{uria2024efficient}, and present a comprehensive study of the magnetic ground and excitonic states of CrCl$_{\rm 3}$ across the strain-induced transition. From our ab initio BSE calculations, when the FM magnetic ordering is changed to the AFM one, we observe the modification of the band structure (as reflected by the appearance of Kramers degeneracy) and a small band gap increase from  3.89 to 4.04 eV. Of particular interest is the identification of the crucial modification of optical properties and the significant redshift of the lowest exciton when going through the FM-AFM transition. The structure of the paper is as follows. In Section II, we briefly introduce the simulation methods whereas section III is devoted to the numerical results of the study, focusing on the electronic and optical properties of the CrCl$_{\rm 3}$ crystal structure. Finally, we summarize our results in section IV.

\section{Theoretical and Computational Methods}\label{sec:theory}
The electronic and linear optical properties of semiconductor crystal structures are calculated in the framework of DFT and many-body perturbation theory. To determine the ground-state properties of CrCl$_3$, we perform DFT calculations with the QUANTUM ESPRESSO package \cite{quantum}, using the range-separated hybrid scheme (HSE06) \cite{hybrid1,hybrid2} modification of a PBE-type \cite{gga} generalized gradient approximation functional. In the HSE06 method, the short-ranged exchange is constructed by mixing 25 $\%$ exact nonlocal Hartree-Fock exchange and 75 $\%$ semi-local PBE exchange. We use norm-conserving pseudopotentials with a plane- wave energy cutoff of 60 Ry. Convergence tests have been performed carefully both for plane-wave cutoff energy and k-point sampling. A $12\times12\times1$ Monkhorst pack k-point mesh was used in the computations. All the lattice constants and atomic coordinates are optimized until the maximum force on all atoms is less than $10^{-2}$  $eV$/${\\A}$. A large vacuum space of at least 20 ${\\A}$ thick is included in the unit cell to avoid interaction between images. After the self-consistent calculation, we perform a Wannier interpolation using the WANNIER90 package \cite{pizzi2020wannier90, mostofi2008wannier90}.

The optical spectra are obtained by solving the BSE; the equation of motion of the two-particle Green's function ~\cite{Hanke, Strinati} in the Tamm-Dancoff approximation. The matrix eigenvalue form of the BSE is given by~\cite{Onida, Rohlfing, ebrahimian2021natural}
\begin{equation}\label{n}
\sum_{\rm \nu^{\prime} c^{\prime}\bm k^{\prime}} {\cal H}^{{\text {BSE}}}_{\bf \nu ck,\nu^{\prime} c^{\prime}k^{\prime}}{A}^{\lambda}_{\nu^{\prime} c^{\prime}{\bf k}^{\prime}} = {E}^{\lambda}{A}^{\lambda}_{\bf \nu ck}
\end{equation}
where $\nu$, $c$ and $\bm k$ indicate the valence band, conduction band crystal momentum in the reciprocal space, respectively. Eigenvalues ${E}^{\lambda}$ and eigenvectors ${A}^{\lambda}_{\bf \nu ck}$ represent the excitation energy of the $\lambda$th-correlated electron-hole pair and the coupling coefficients used to construct the exciton wavefunction, respectively.
To obtain the exciton energies and wavefunctions, we use the XATU-Wannier implementation \cite{uria2024efficient} of the BSE,  where orbitals are treated as point-like and an effective Rytova-Keldysh screened interaction is considered. The screening length is obtained as explained in Ref. \cite{Cudazzo.PhysRevB.84.085406}.

\begin{figure}[t]
	\centering
	\includegraphics[width=1.0\linewidth]{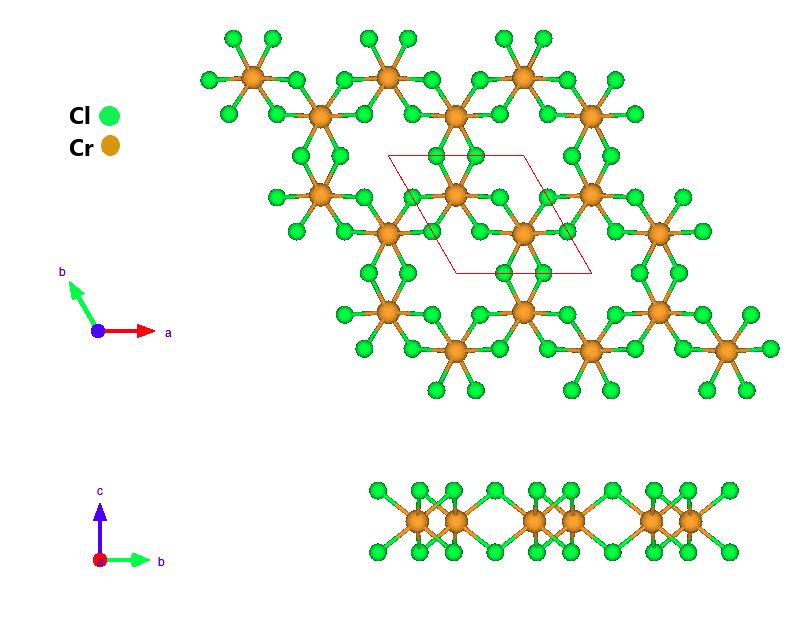}\\
	\caption{(Color online) Top view (upper panel) and side view (lower panel) of the crystal structure of monolayer CrCl$_3$. 
		\label{fig:1}}
\end{figure}

\section{\label{sec:level3} Numerical Results and Discussions}
The transition chromium trihalides CrX$_3$ (X = Cl, Br, I) represent a family of layered magnetic van der Waals materials that exhibit a rhombohedral structure with the R-3 symmetry at low temperatures, and a monoclinic structure with C2/m symmetry at high temperatures. The Curie temperature for the layered ferromagnetic CrCl$_3$ is $27$ K and the ferromagnetic state is preserved down to the single-layer limit \cite{huang2017layer}.A schematic plot of the monolayer CrCl$_3$ is shown in Fig.~\ref{fig:1}. The CrCl$_3$ monolayer consists of Cl-Cr-Cl triple layers where the magnetic Cr atoms are arranged in the honeycomb structure and are surrounded by six Cl atoms according to the octet rule.
 
\begin{figure*}[t]
\centering
\includegraphics[width=0.88 \textwidth]{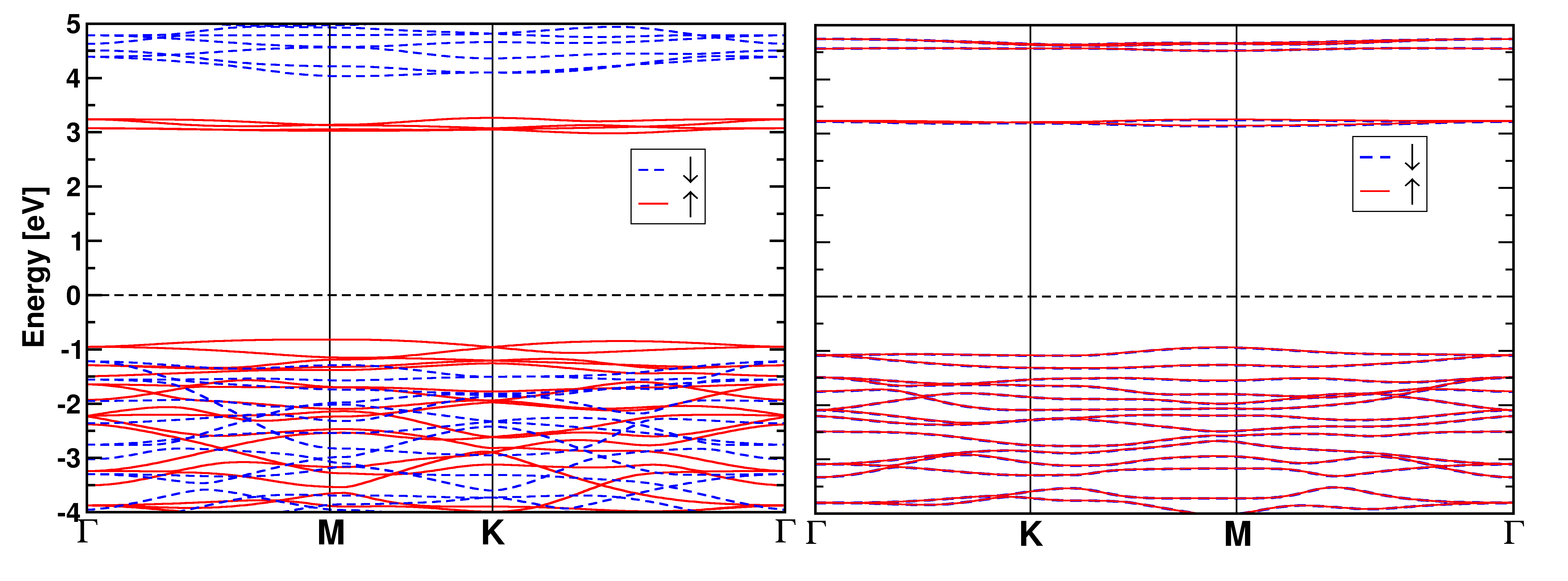} \\
\caption{(Color online) The HSE06 calculated electronic band structure of monolayer $CrCl_3$ in ferromagnetic (left panel) and antiferromagnetic (right panel) phases. For the ferromagnetic phase, the red (blue) color denotes the major-spin (minor-spin) polarization. For the antiferromagnetic phase, the bands are spin degenerate, with opposite spin components from each Cr atom in the unit cell. There is a small band gap increase when the phase changes from being ferromagnetic to antiferromagnetic (3.89 versus 4.06 eV).
\label{fig:2}}
\end{figure*}

In the single-layer limit, CrCl$_3$ retains its ferromagnetism and it is shown by experiments to have an in-plane magnetic easy axis.
The partially filled $d$ orbitals of Cr$^{3+}$ ion forms a $3d^3$ electronic configuration which is highly affected by the Mott-Hubbard mechanism inducing magnetism and a band gap in the electronic structure. In fact, due to electric crystal fields of octahedral arrangement, the $d$ orbitals of the Cr atoms decomposed into a set of triply degenerate orbitals t$_{2g}$ (with lower energy) and doubly degenerate orbitals e$_g$ (with higher energy). Minimizing the orbital energy and Coulomb interaction energy, each of the three t$_{2g}$ orbitals is occupied by one electron which result in an atomic magnetic moment of 3$\mu_{B}$.%

\begin{table}[b]
\caption{\label{tab:table1}
Convergence test of HSE band gaps and A to E exciton energies for FM and AFM monolayer CrCl$_3$. The unit is $eV$.}
\begin{ruledtabular}
\begin{tabular}{cccccccc}
 CrCl$_3$ &A &B &C &D &E (Dark) & HSE band gap\\
\hline
FM& 0.63 & 0.89 & 1.44 
& 1.84 & 0.59 & 3.89 \\
AFM& 1.01 & 1.22 & 1.36 
& 1.73 & 1.32 & 4.04 \\
  
\end{tabular}
\end{ruledtabular}
\end{table}

Starting from an experimental reference, we performed structure optimization for both magnetic phases.  Our results show that the difference in the total energy ($E_{\rm FM}-E_{\rm AFM}$) is -11.4 meV, confirming the FM phase for the ground state of the monolayer.
The softness of the monolayer suggests lattice strain as an adjustable parameters for tuning and engineering of the electronic structure and magnetic ordering in this system. Thus, we  study the structural modification and related impact on magnetism and optical response resulting from biaxial strain. To explore the impact of strain on the magnetic ground state, we apply an in-plane biaxial strain, defined as $\epsilon =(\textbf{a}-\textbf{a}_0)/\textbf{a}_0 $, where $\textbf{a}_0$ and \textbf{a} are lattice parameters in the absence and presence of a strain field, respectively.  The energy difference between the two magnetic orderings, namely, FM and AFM  under different strains has been later calculated. The calculations show that by applying a compressive biaxial strain field a quantum phase transition from FM to AFM happens around $- 2.5 \% $ . The AFM state remains more stable than the FM configuration with increasing compressive strain field. This is in a good agreement with previous theoretical works \cite{webster2018strain, ebrahimian2023control}.

The HSE electronic band structures for the FM and AFM phases are shown in Fig~\ref{fig:2}.  A small increase in the band gap is observed from the FM to AFM phase (upon compression) in this system. For the FM configuration, all bands are spin polarized with a 3.89 eV energy gap between the valence band maximum (VBM), which is located at the high-symmetry M point, and the conduction band minimum (CBM), located around the middle point of the K-$\Gamma$ line. On the other hand, the AFM magnetic phase (Fig.~\ref{fig:2}) has an energy gap of 4.04 eV at the M point. In this case, the bands are Kramers degenerate due to the combined effect of parity and time-reversal (neither of them symmetries on their own). It is worth noticing that for both FM and AFM orderings, valence and conduction band edges exhibit nearly flat dispersions. This enhanced joint density of states indicates strong many-electron interactions and potentially active inter-band transitions due to the spin-allowed selection rule. This behavior is even more pronounced in the AFM configuration.

\begin{figure}
	\includegraphics[width=1.0\linewidth]{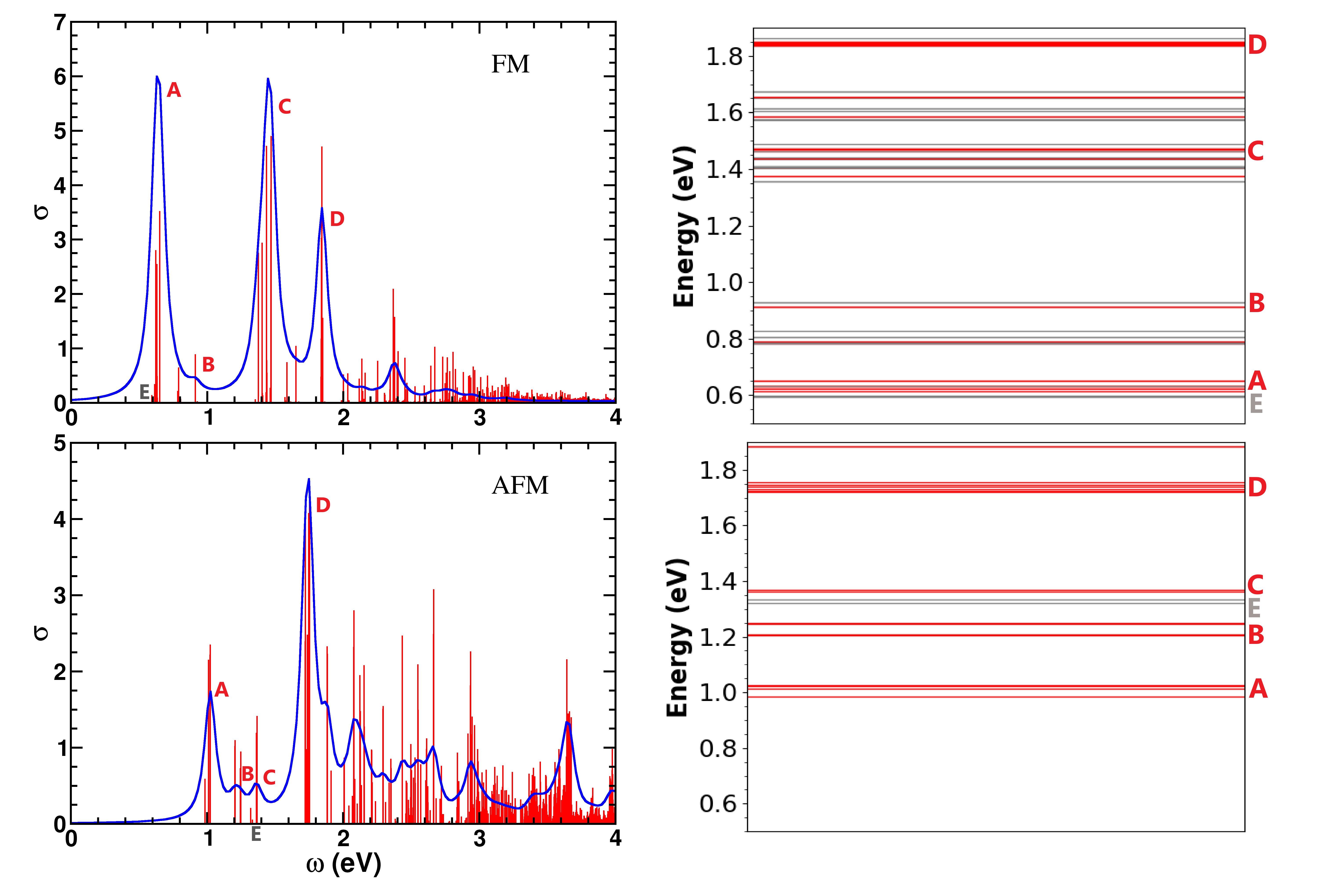}\\
	\caption{(Color online) Left panel: Optical conductivity of monolayer CrCl$_3$ for in-plane polarized light with electron-hole interaction (HSE-BSE) in FM (top panel) and AFM (lower panel) order. The solid blue line is the spectrum by BSE while the vertical (red) bars represent the relative oscillator strengths for the optical transitions. Right panel: Exciton energy levels of monolayer CrCl$_3$ in FM (top panel) and AFM (lower panel). Optically bright exciton states are in red while dark ones in gray. We label the bound exciton states with E for the  dark states and A-D for the bright states as evident in the plot of BSE spectrum.
\label{fig:op}}
\end{figure}

We have also calculated the optical conductivity spectrum of both magnetic phases, including excitonic effects. As seen in Fig.~\ref{fig:op}, the calculated spectrum of the FM configuration features four peaks at around 0.63, 0.89, 1.44 and 1.84 eV, which are composed of several bright excitonic states in each peak and are denoted by A, B, C and D, respectively. These low-energy excitons are attributed to the spin-allowed transitions within the majority spin channel only. Notably, the C peak agrees with recent photoluminescence measurements of monolayer and multilayer CrCl$_3$ at 2 K, exhibiting a single peak around 1.43 eV \cite{cai2019atomically}. The A and B excitons are also observed experimentally and have been predicted theoretically\cite{acharya2022real, dillon1966magneto}. It is evident from our calculation (Fig.~\ref{fig:op}) that there are also several dark states around these prominent peaks. Specifically, as for CrI$_3$\cite{wu2019physical}, due to the existence of two Cr atoms in a unit cell, there exist two dark states (excitons E) with nearly the same energy at the onset of  spectrum. For the optical conductivity spectrum of the AFM configuration, the four lowest energy peaks (denoted as A, B, C and E respectively) are located around 1.01, 1.22, 1.36 and 1.73 eV and composed of several bright exciton states in each peak. Here the first two dark states (excitons E) are found at higher energy around 1.32 eV so their binding energy are smaller than the corresponding ones in the FM phase.

To compare the excitons in the CrCl$_3$ monolayer of different magnetic orders, we focus on the optical response at the low energy regime below 2 eV. In both FM and AFM phases, the peak of lowest energy in the spectrum is followed by a small peak. These have lower energies at the FM configuration; the excitons A and B are shifted down in energy by about 0.3 eV from the AFM to FM configurations. Notice that this shift is larger than the bare single-particle band gap reduction.
In the same manner, the first dark exciton also moves towards a lower energy by a factor of nearly 2 from the AFM to the FM phase and becomes the onset of the spectrum. In contrast, we observe that excitonic peaks C and D have higher energies in the FM  than in the AFM phase.

\begin{figure}
	\includegraphics[width=1.0\linewidth]{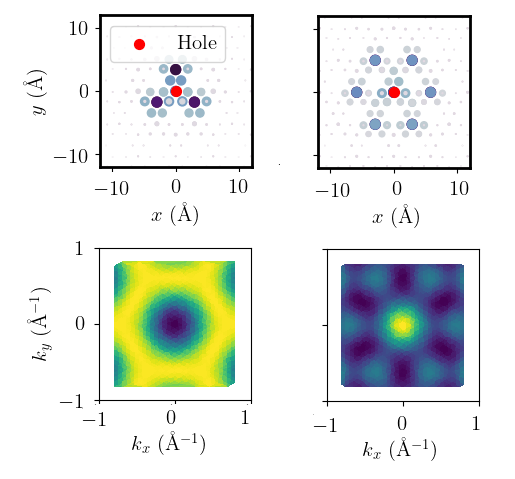}\\
	\caption{(Color online) Real (top panel) and reciprocal-space (lower panel) probability densities of the first exciton energy levels (A) in  FM (left panel) and AFM (right panel) configurations of monolayer CrCl$_3$ with energy of 0.63 and 1.01 eV,  respectively. The red dot in the real-space exciton probability densities shows the position of the hole which is fixed on a Cr atom. As shown, the A excitons  are delocalized (localized) in reciprocal space for FM (AFM) and strongly localized (delocalized) in real space.
\label{fig:ex}}
\end{figure}

To further explore the nature of the excitons we plot the real-space distributions of the excitons with different magnetic orders where the hole is positioned on the Cr atom mainly involved in the construction of the valence band near the Fermi level (Fig.~\ref{fig:ex}). Because of the large e-h binding energy (3.26 eV), bright A excitons in the FM phase have highly localized wave functions distributed around the nearest Cr atoms and barely extending over one unit primitive cell. By contrast, A bright excitons in the AFM phase show Wannier behavior, extending over several unit cells with a larger exciton radius in real-space and confirming that the binding energy of this exciton is reduced (3.03 eV). This explains the increase in excitons energy position across the strain induced spin flip transition from FM to AFM.

Breaking the time reversal symmetry in the FM configuration leads to band splitting, causing the appearance of four conduction bands near the Fermi level. In the FM phase, these four majority spin up conduction bands are actively involved in the formation of the A excitons  with the lowest excitation energy. Since the spin of all atoms points in the same direction, nearest-neighbor atoms are accesible in the excitonic transitions, favoring a small exciton size and strong binding. On the other hand, our calculations assign the lowest dominant optical A resonance in the AFM phase to transitions from the valence to the two lowest conduction degenerate bands. 
In the AFM configuration, both excitation spin channels are degenerate and are allowed. However, spin conservation favors the electrons from each Cr atom to hop onto next-nearest same spin Cr atoms, while hopping to nearest neighbor opposite spin Cr atoms is hindered. As a result, the e-h wave functions extend over several unit cells reducing the e-h interactions, which is reflected in the increase of the excitons energy position in the AFM phase.

\begin{figure*}[t]
\centering
\includegraphics[width=0.98 \textwidth]{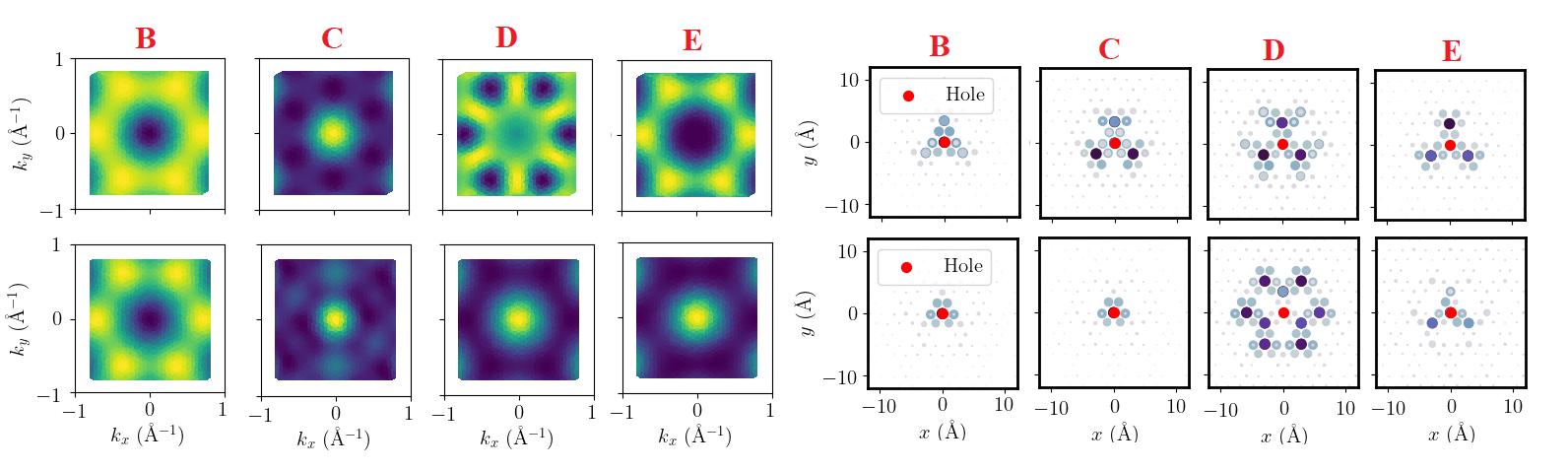} \\
\caption{(Color online)  Left panel: Reciprocal (left panel) and real(right panel)-space probability densities of the B, C, D, E excitons in FM (top panel) and AFM (lower panel) configurations of monolayer $CrCl_3$. The red dot in the real-space exciton probability densities plot shows the position of the hole which is ﬁxed on a Cr atom. 
   \label{fig:ex2}}
\end{figure*}

The \textbf{k} probability densities of the A excitonic states are also shown in Fig.~\ref{fig:ex}. For A excitons in the AFM phase, wavefunctions are localized around the center of Brillouin zone which means that the related transitions take place around the $\Gamma$ point, mostly between two nearest valence and conduction bands close to the Fermi level. 
As seen in Fig.~\ref{fig:ex}, for FM order, A excitons spread over the \textbf{M}- \textbf{K} path mostly contributed by the excitations originate from the upper valence bands to the first four conduction bands. This plot is consistent with the intuition that delocalized excitons in real space result in strongly localized excitons in reciprocal space. As the valence and conduction bands near the Fermi level in both both FM and AFM phases are mostly contributed by Cl-p and Cr-d orbitals, the A excitons result from the p$\rightarrow$d transition. In other words, the d-p hybridization between Cr and Cl states is the symmetry breaking mechanism needed to allow for these transitions. 

The real-space wave functions and  \textbf{k} probability densities of bright excitons with higher energies and dark E excitons are depicted in Fig. ~\ref{fig:ex2}. As shown in this figure, the wavefunctions of B excitons in FM and AFM phases are the same as A excitons except that the transitions for these excitons take place around the K point in the AFM phase. The C and D excitons in the AFM configuration are localized around the center of the Brillouin zone, meaning that the related transitions take place around the $\Gamma$ point (Fig. ~\ref{fig:ex2}). In the FM phase, the wavefunctions of the C excitons are localized around the center of the Brillouin zone, while the D excitons take place around the M point.
As shown in Fig. ~\ref{fig:ex2}, the wave functions of C and D excitons in FM phase are significantly delocalized in nature compared to the other extreme of the AFM phase and extend over several unit cells due to the smaller exciton binding energy. This is consistent with the higher energy position of the B and C excitons of the FM phase when compared to those of the AFM configuration.  The dark E excitons in AFM phases are localized around the center of Brillouin zone while they are  around the K point in FM phase.
Figure ~\ref{fig:op} shows that the E excitons in the FM phase (AFM) are close to the A peak (C) at 0.59 (1.32) eV. This means that they move from 0.59 to 1.32 eV across the transition from the FM to the AFM phases, which result in smaller binding energy for dark E excitons in the AFM phase. This explains that the real-space distributions of the E excitons in AFM phase are more  delocalized.

\section{Summary and concluding remarks}
In conclusion, we have investigated the magnetic properties and exciton transitions of the monolayer CrCl 3 under biaxial strain from ab initio many-body perturbation theory calculations.
By applying a compressive biaxial strain field we observe a quantum phase transition from FM to AFM happens around $- 2.5 \% $.
In both phases, the observed electron-hole interactions  in the optical responses of the monolayer CrCl$_3$ are strongly influenced by magnetic ordering.
When the magnetic ordering of AFM is changed to FM we observe the crucial modification of optical properties and the significant redshift of the lowest excitons.
First, bright excitons in the FM phase have a highly localized wave function distributed around the nearest Cr atoms and nearly extending over one unit primitive cell, while in the AFM it shows Wannier behavior, extending over several unit cells with a larger exciton radius in real space. This explains the increase in exciton energy positions across the strain-induced spin-flip transition from the FM to the AFM phase. These results open up a practical approach to tune the intralayer mutual interactions between spins and excitons  in 2D magnetic semiconductors through mechanical means. 

\section{Acknowledgments}
We acknowledge financial support from the Spanish Ministry of Science and Innovation through Grants 
TED2021-131323B-I00 and PID2022-141712NB-C21, the María de Maeztu Program for Units of Excellence in R\&D (Grant CEX2023-001316-M), the project ``Disruptive 2D materials'' (MAD2D-CM-UAM7), funded by Comunidad de Madrid within the Recovery, Transformation and Resilience Plan, and by NextGenerationEU programme from the European Union. We also acknowledge computational resources from CCC at UAM and Spanish RES.

\nocite{apsrev41Control}
\bibliographystyle{apsrev4-1}
\bibliography{H}

\end{document}